\documentstyle[12pt]{article}
\input epsf
\begin{document}

%
\def\figevolve
{
\begin{figure}[tbp]
\begin{center}
\begin{minipage}[h]{6.5in}
\epsfxsize=6.3in
\epsfbox[36 144 520 650]{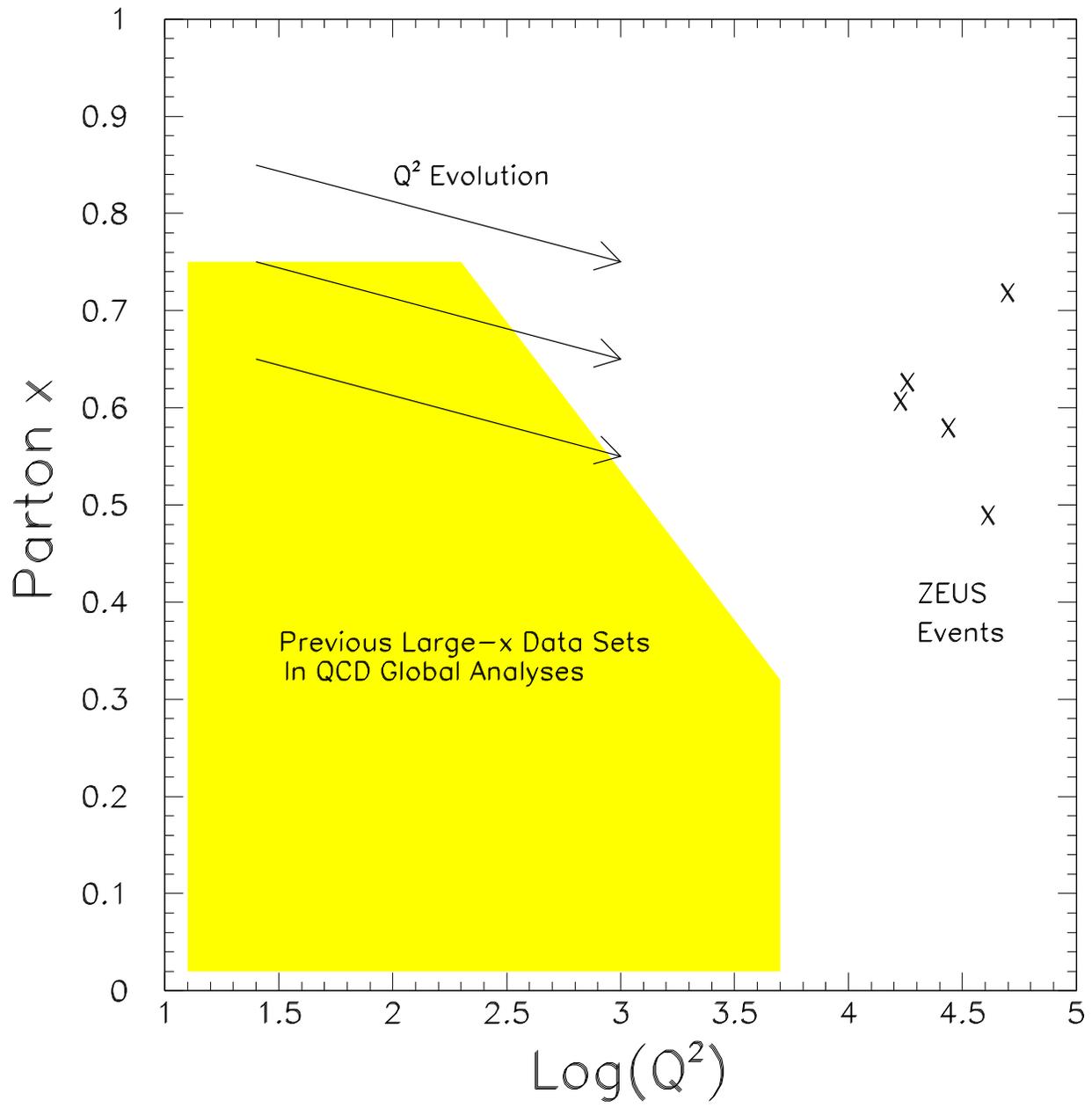}
\caption{How large-$x$ quarks can evolve into the HERA kinematic
region.}
\label{evolve}
\end{minipage}
\end{center}
\end{figure}
}
\def\figparam
{
\begin{figure}[tbp]
\begin{center}
\begin{minipage}[h]{6.5in}
\epsfxsize=6.3in
\epsfbox[36 144 520 650]{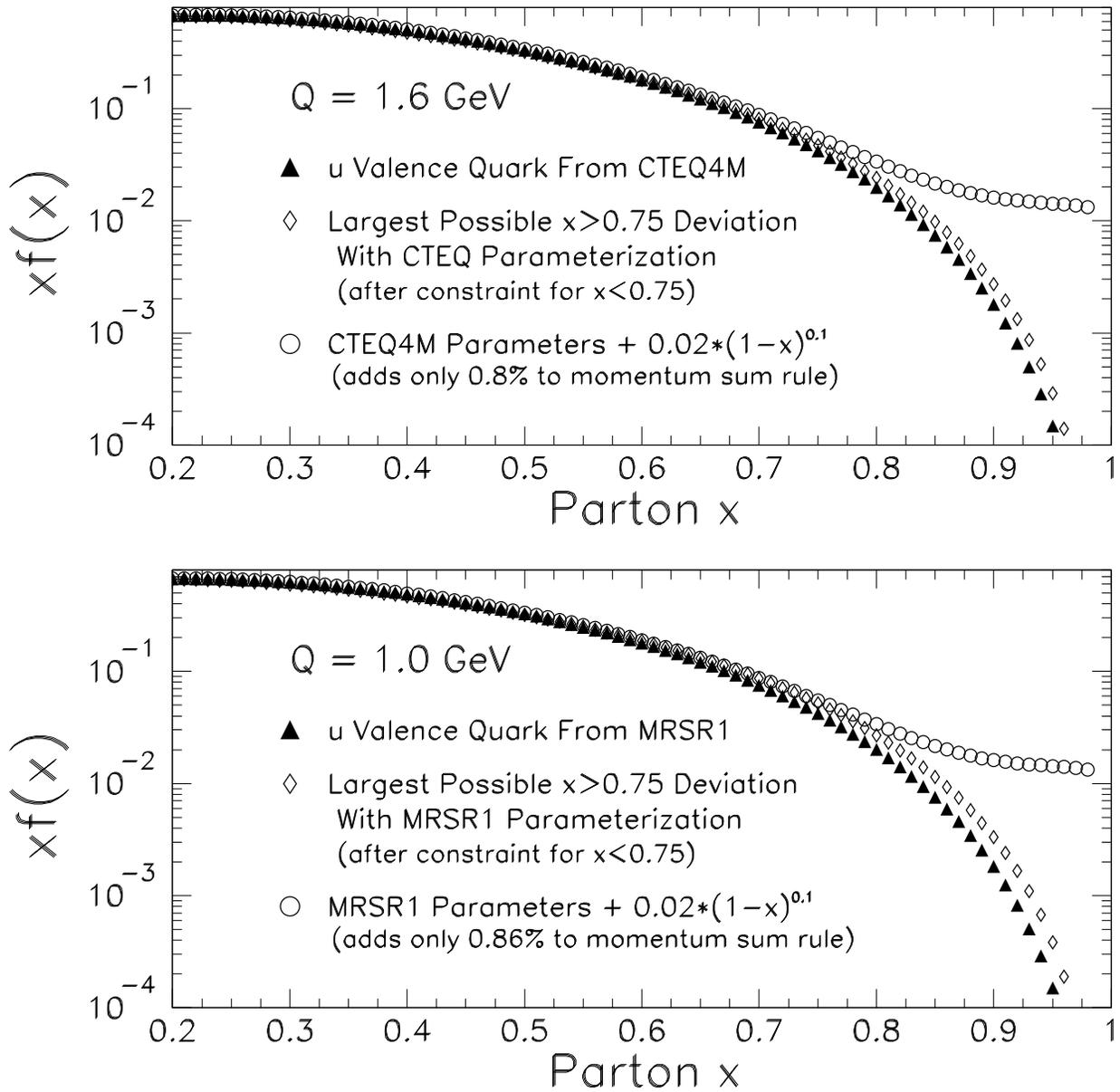}
\caption{The u valence distributions of CTEQ4M and MRSR1 are
compared to variations described in the text.}
\label{param}
\end{minipage}
\end{center}
\end{figure}
}
\def\figuvadd
{
\begin{figure}[tbp]
\begin{center}
\begin{minipage}[h]{6.5in}
\epsfxsize=6.3in
\epsfbox[36 144 520 650]{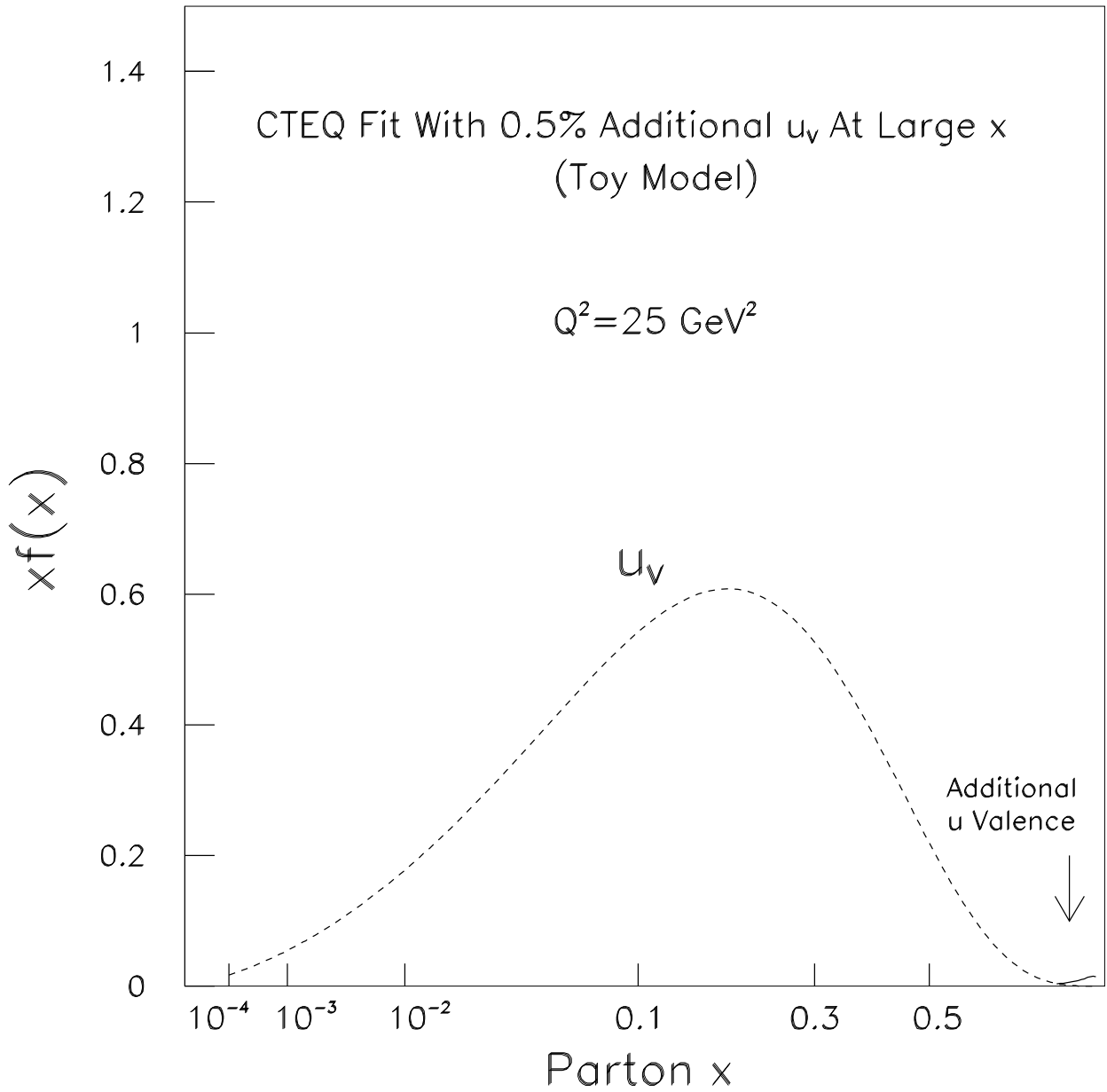}
\caption{The u valence distribution from the
CTEQ4M fit (dashed line) and the additional 0.5\% component.}
\label{uvadd}
\end{minipage}
\end{center}
\end{figure}
}
\def\figtotq
{
\begin{figure}[tbp]
\begin{center}
\begin{minipage}[h]{6.5in}
\epsfxsize=6.3in
\epsfbox[36 144 520 650]{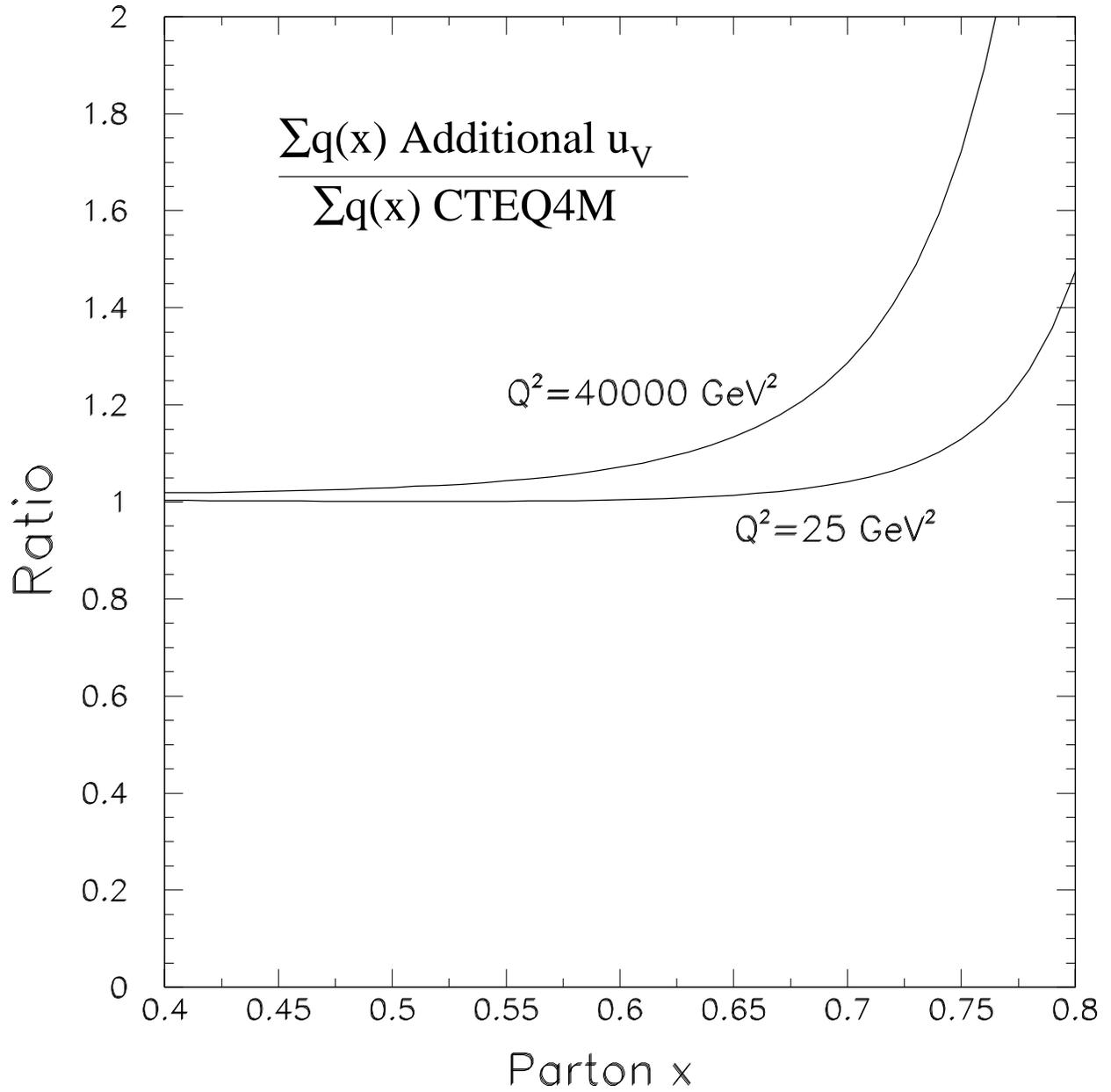}
	\caption{Comparison of the sum of the quark distributions
from a conventional analysis, and with the fit including a 0.5\%
additional component of $u_v$.}
	\label{totqrat}
\end{minipage}
\end{center}
\end{figure}
}
\def\figbcd
{
\begin{figure}[tbp]
\begin{center}
\begin{minipage}[h]{6.5in}
\epsfxsize=6.3in
\epsfbox[36 144 520 650]{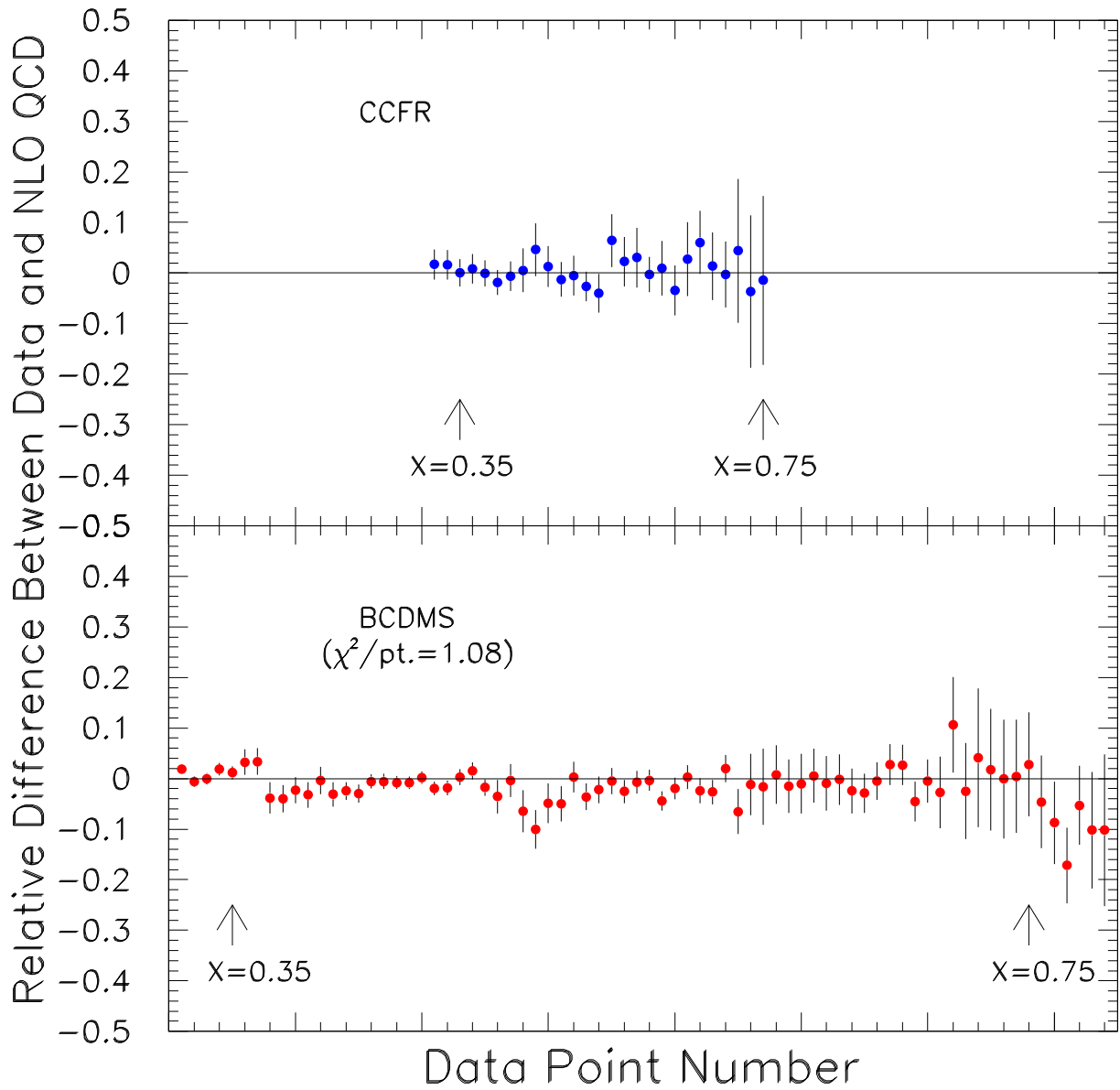}
	\caption{The BCDMS and CCFR data sets are shown compared to
NLO QCD calculations using the fit including the 0.5\% additional
component of u valence quarks. The agreement is
good in both cases. (see text)}
	\label{bcdms}
\end{minipage}
\end{center}
\end{figure}
}
\def\figjet
{
\begin{figure}[tbp]
\begin{center}
\begin{minipage}[h]{6.5in}
\epsfxsize=6.3in
\epsfbox[36 144 520 650]{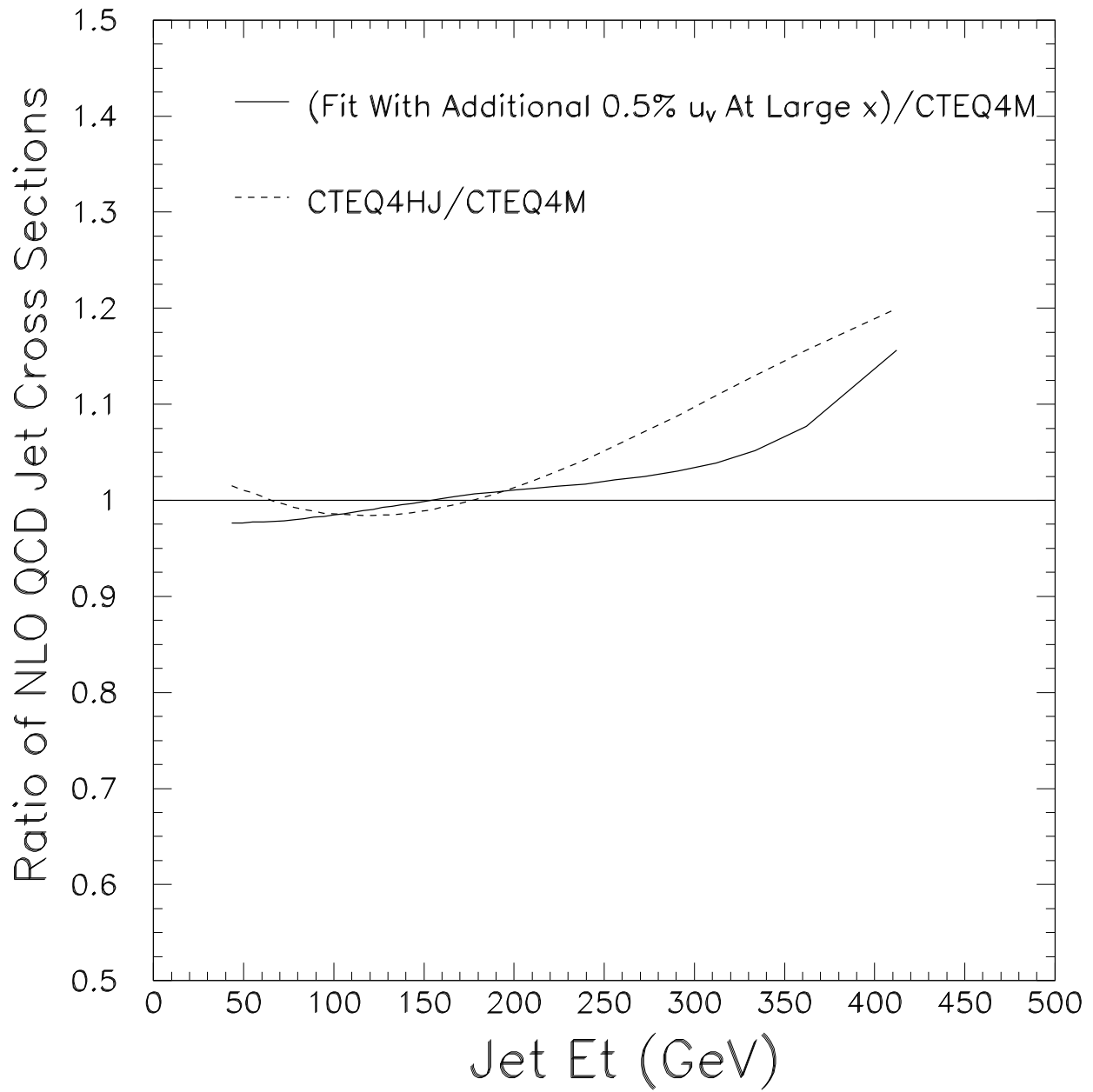}
	\caption{Comparisons of NLO QCD jet cross sections
from a conventional analysis (CTEQ4M), an analysis with
additional high-$x$ gluons (CTEQ4HJ), and an analysis
with an additional 0.5\% component of $u_v$ at large-$x$.}
	\label{jet}
\end{minipage}
\end{center}
\end{figure}
}
\def\figbcdeu
{
\begin{figure}[tbp]
\begin{center}
\begin{minipage}[h]{6.5in}
\epsfxsize=6.3in
\epsfbox[36 144 520 650]{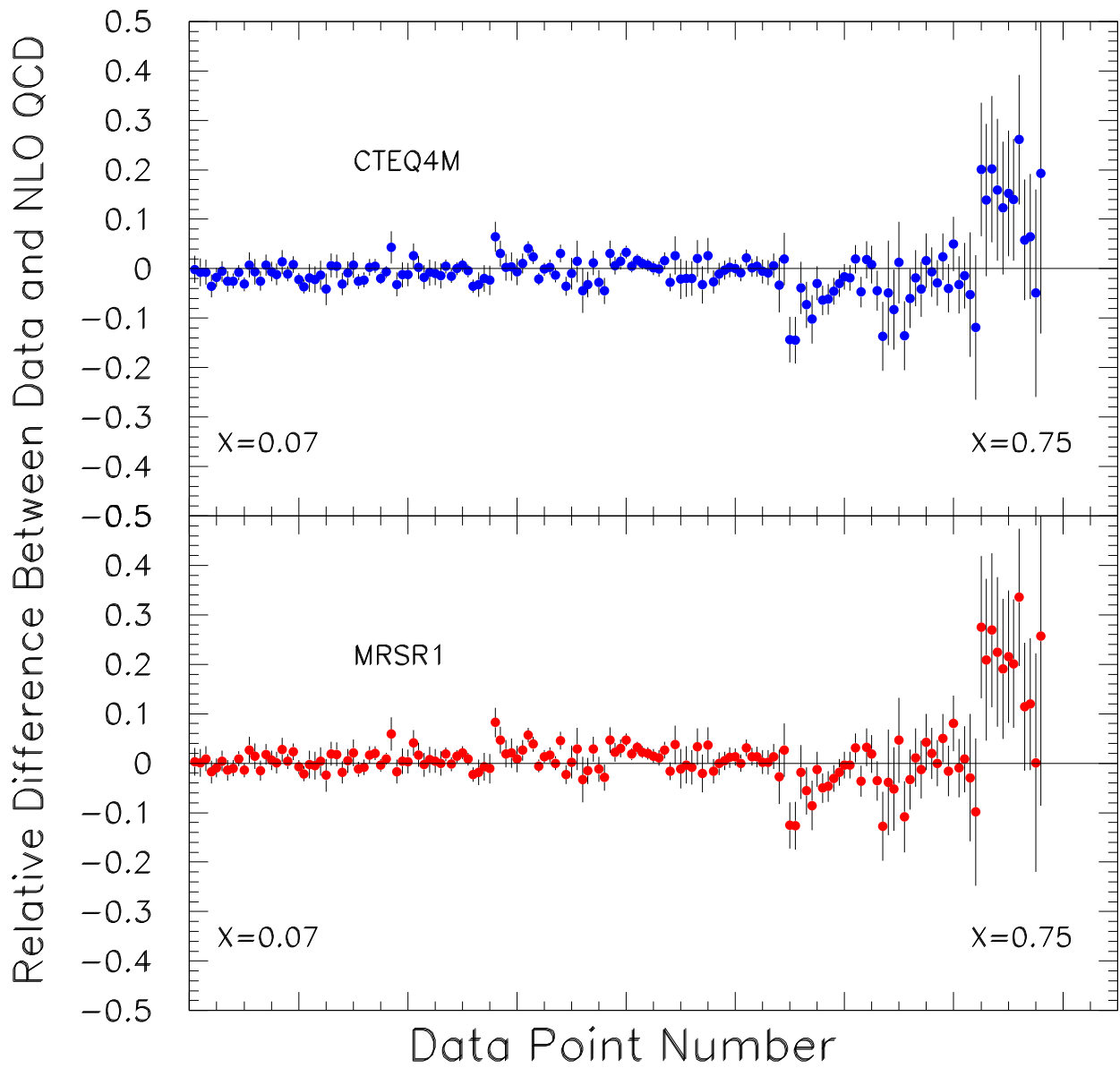}
	\caption{The BCDMS data with a deuterium target
is shown compared to NLO QCD with the CTEQ4M partons and
with the MRSR1 partons.}
	\label{bcdeu}
\end{minipage}
\end{center}
\end{figure}
}
\def\figev2
{
\begin{figure}[tbp]
\begin{center}
\begin{minipage}[h]{6.5in}
\epsfxsize=6.3in
\epsfbox[36 144 520 650]{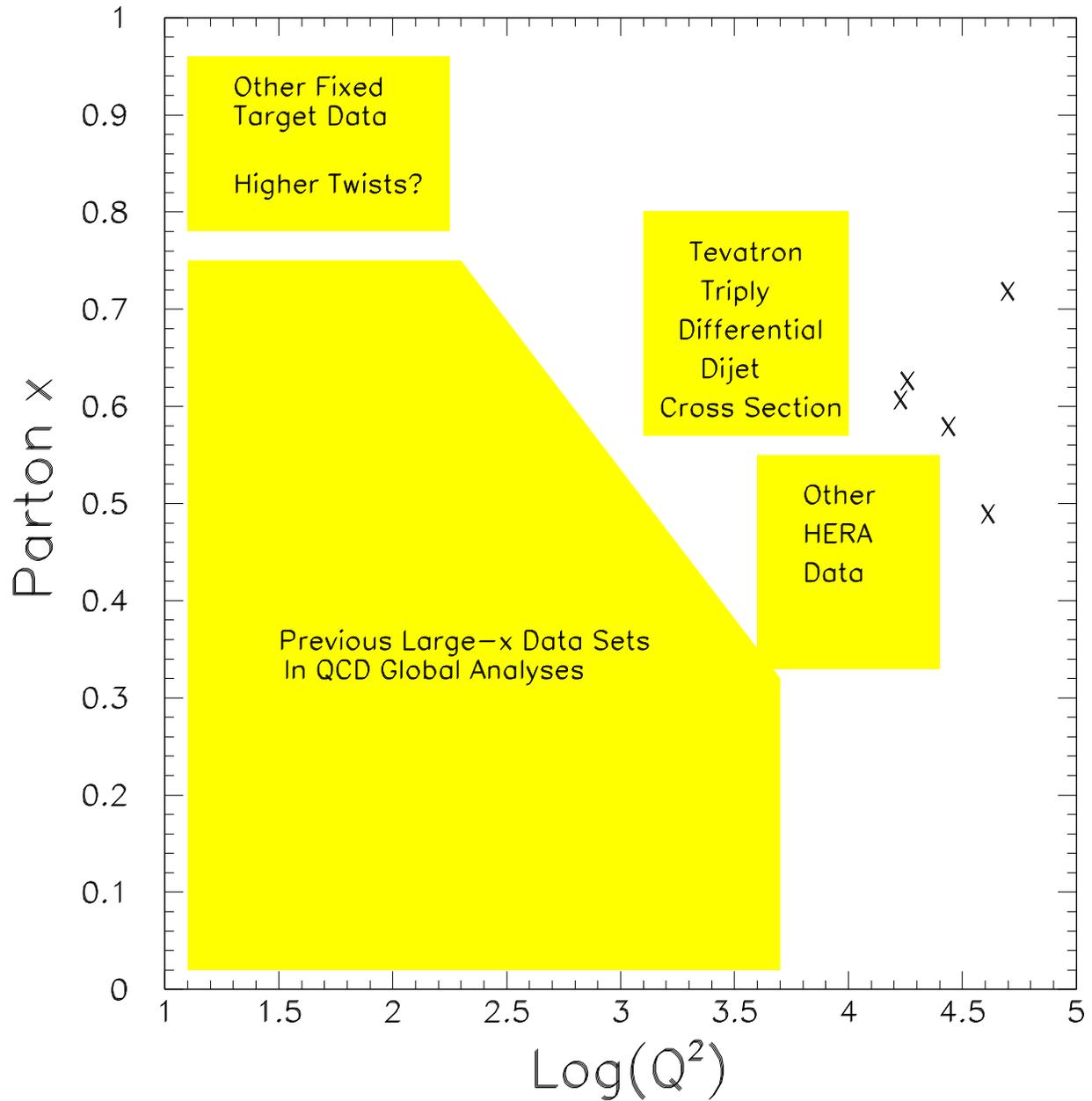}
	\caption{The kinematic regions are shown of three additional
classes of measurements that may provide additional constraints
on the large-$x$ quarks in the future.  The Tevatron dijet and ``Other HERA'' 
blocks are separated only for the purpose of clear labelling; they overlap
throughout the region bridging the established data and the ``ZEUS Events'' 
regions.}
	\label{evolve2}
\end{minipage}
\end{center}
\end{figure}
}
\begin{titlepage}
\begin{tabular}{l}
May 1997                   
\end{tabular}
    \hfill
\begin{tabular}{l}
MSU-HEP-70316 \\
CTEQ-705 \\
\end{tabular}

\vspace{2cm}

\begin{center}

\renewcommand{\thefootnote}{\fnsymbol{footnote}}

{\LARGE HERA Events, Tevatron Jets and \\ Uncertainties on Quarks at Large
x\footnote[1]{
This work was supported in part by the DOE and NSF.}}

\renewcommand{\thefootnote}{\arabic{footnote}}

\vspace{1.25cm}


{\large S.~Kuhlmann$^a$, H.~L.~Lai$^b$, and W.~K.~Tung$^b$}

\vspace{1.25cm}

$^a$Argonne National Laboratory, \\
$^b$Michigan State University

\end{center}
\vfill

\begin{abstract}
The recently reported excess of events at HERA compared to QCD calculations
impels us to examine all possible Standard Model explanations before invoking
``new physics''.  We explore the possibility of adding an unusual, but small,
component of additional quarks at large $x$ (beyond $x>0.75$) as a way to
increase the predicted SM cross-section in the HERA kinematic region by the QCD
evolution feed-down effect.  We describe various scenarios under which
this can be achieved while maintaining
good global fits to all established data sets.  This implies a much 
larger SM uncertainty than commonly assumed. 
In addition, the modified parton
distributions provide another possible mechanism to account for the CDF
high-$p_t$ jet excess which occurs at similar $x$ and $Q^2$ values.
\end{abstract}

\vfill
\newpage
\end{titlepage}

\renewcommand{\baselinestretch}{1.25}

Recently the H1 and ZEUS experiments at HERA~\cite{hera} have reported an
excess of large-$x,Q^2$ deep inelastic scattering (DIS) events compared
to next-to-leading order QCD expectations (NLO QCD). In order to determine
whether this enhancement constitutes a signal for {\it new physics}, it is
crucial to investigate all possible explanations within the Standard
Model. Both experiments attribute a $7-8\%$ uncertainty to the QCD
expectations, using standard procedures based on conventional parton
distributions, which is small compared to the observed effect. In this note we
investigate more closely an important part of this uncertainty -- that relating
to the quark distributions at large $x,Q^2$ beyond the range which is well
measured by existing fixed target experiments. The goal is to explore possible
unconventional features in this uncharted kinematic region and to quantify
their effects on existing experimental data used in global QCD analyses and on
the new observations.

It is useful to carefully define the question we are addressing.  The H1 data
shows a fairly discrete jump in its last $x$ bin which, if correct, certainly
rules out a parton distribution interpretation, since QCD effects at such large
$Q^2$ should be smooth. 
If the excess persists it is still an open question whether, with increased statistics, the
excess will be more spread out like the
present ZEUS data.  This is the scenario that our study addresses.  We shall ask
the question: under what conditions, if any, is it possible to have a
substantially increased quark contribution to the DIS cross-section in the new
HERA kinematic range and still maintain consistency with all established lower
energy data?  We shall not be concerned with describing any detailed features
of the new HERA data, nor include them in the analysis directly.

The first task is to try to improve the existing estimates of the uncertainties
on $q(x,Q^2)$ due to a conventional parton distribution analysis.  We performed
a series of special purpose global fits designed to quantify how far $q(x,Q^2)$
can be forced to deviate from its nominal values at the kinematic region of the
HERA events and still maintain a good fit to the existing global data sets.
Using conventional functional forms for the input non-perturbative parton
distributions such as those used by the CTEQ \cite{cteq4} and MRS \cite{mrs}
groups, we confirm the H1 and ZEUS results: one can induce at most a $\approx
7\%$ change in parton distributions in the relevant kinematic range.

Therefore to find a way to produce a larger change in the quark distributions,
one must go beyond the usual assumptions. Because of the well-known feed-down
effect of QCD evolution (as the result of parton splitting with increasing
resolution power), the observed effect in the range $0.5<x<0.7$ and $100$
GeV$<Q<200$ GeV can be obtained only by introducing some unexpected large $x$
feature at the low $Q$ value where QCD evolution starts.  In order not to upset
the existing agreement between theory and established fixed-target measurements
which extend to $x\sim 0.65-0.75,$ $Q\sim 20$ GeV, the new feature has to be
fairly close to $x=1$ if it sets in at $Q=Q_0=1-2$ GeV.\footnote{If $Q_0$ is
higher, say $5$ GeV for a b quark, then the new feature can be more spread out
in $x$.}  This is graphically illustrated in figure~\ref{evolve}. This graph
also suggests perhaps the feed-down mechanism is the only possible one with the desired effect involving parton distributions.

Since very little experimental information is available about the flavor dependence of parton distributions above $x=0.75$, one can in principle explore a variety of
possibilities: variations in the valence quarks $u_v,\ d_v$ as well as unusual
non-perturbative heavy flavor components. We begin by examining the $u_v$
distribution which gives the largest contribution to the cross-section.  To
demonstrate how restrictive the commonly used parameterizations for valence
quark distributions are in regard to changes in the $x>0.75$ region, the following exercise is performed: we
randomly varied all 5 parameters in the $u_v$ distribution (and all
combinations), and removed trials that changed the total momentum of the $u_v$
distribution by more than 3\%.  We also removed trials that disagreed with the
conventional distribution by more than 5\% between $x=0.01$ and $x=0.65$.  Of
the remaining trials we found the one that gave the largest increase in the
distribution for $x>0.75$, this is shown in Fig.~\ref{param} (as diamonds) for
both the CTEQ and MRS parameterizations.  Compared to the original
CTEQ4M~\cite{cteq4} and MRSR1~\cite{mrs} distributions (dark triangles), the
maximum deviation is very small.  On the other hand, if a small
non-conventional component of the form $0.02*(1-x)^{0.1}$ is added to the
original parameterizations with the same constraints mentioned above at lower
$x$, one obtains a dramatically increased valence distribution for $x>0.75$
(shown by the circles in Fig.~\ref{param}).  This modification is somewhat too
large to agree with the fixed target data, but the exercise demonstrates that
to explore the possibility of increasing the quarks in the HERA kinematic
region, it is necessary to go beyond the conventional parameterizations.

We will now systematically investigate what are the effects of introducing a
small addition of $u_v$ quarks near $x=1$ in the global analysis. Since we are
only concerned about the possibility of establishing a generic scenario, rather
than trying to advance a realistic model, we shall implement this additional
component in the form of a simple function concentrated near $x=1$ at $Q_0=1.6$
GeV. The optimal amount of this component, measured in terms of the momentum
fraction carried, will turn out to be in the range $0.5-1\%.$ The exact shape
is not important as long as the initial shape is sharply peaked, as we are
assuming in our simple model. We find it is not hard to obtain a satisfactory
global fit to established data with such an additional component. The resulting
$u_v$ distribution at a typical scale for existing experiments, $Q=5$ GeV, is
shown in figure~\ref{uvadd}, along with the $u_v$ distribution from CTEQ4M.

The additional large-$x$ component of the $u_v$ distribution has a significant
impact on the quark distributions in the new HERA $Q^2$ range. In
figure~\ref{totqrat} we show the ratio of the sum of the quark distributions
with the new component divided by the same sum with CTEQ4M.  Two curves are
shown, one for $Q^2=40,000$ GeV$^2$ and one for $Q^2=25 $ GeV$^2$.  At $x=0.7$
the change in the quarks is 30\%, much larger than the 6\% uncertainty coming
from the conventional analyses.  In the lower $Q^2$ fixed target region, the
change is minimal.  This is because it takes a significant amount of QCD
evolution for the additional component to propagate down to the $x$ values of
the fixed target experiments, and by then the region of $Q^2$ is beyond that
probed at fixed target experiments.  The comparison of this fit to large-$x$
DIS data is shown in figure~\ref{bcdms}. Plotted are the relative differences
between the large $x$ DIS data of the BCDMS~\cite{bcdms} and CCFR \cite{CCFR}
experiments and the theoretical values from this global fit. Although the
effect of the additional $u_v$ component can already be seen in the highest $x$
bin of the BCDMS data, the general agreement is still quite good, and the
agreement with the CCFR data is excellent.  The quality of this new global fit
to other DIS, DY, W-asymmetry and direct photon data sets are indistinguishable
from that of standard parton distributions such as CTEQ4M. Thus, it serves as
an example of a new class of viable parton distributions consistent with
existing data below $x<0.75$, with the feature that it gives rise to a
significantly larger quark distribution than conventional sets at large
$x,Q^2.$

Previously CDF had reported an excess of high-$p_t$ jets compared to
conventional NLO QCD expectations~\cite{cdfjet}. The relevant $x$ and $p_t$
range of their high $p_t$ events ($0.3<x<0.5; \ 200 < p_t < 450$ GeV) are
remarkably similar to that of the HERA events. Since jet production in this
region is mostly due to quark-quark scattering, any modification of the quark
distributions will affect jet cross sections. It has been known that it is
impossible to modify quark distributions of the conventional type to fit the
CDF jets simultaneously with fixed target DIS data~\cite{cteqglu,GMRS}. But
the additional $u_v$ component introduced above does fit fixed target data and
it will give additional contribution to the jet cross section at the high $p_t$
range in the right direction as observed by CDF. To see this effect in
quantitative terms, we compare in Fig.~\ref{jet} the inclusive jet cross
sections calculated using three different sets of parton distributions: (i) the
new distributions, including 0.5\% additional $u_v$; (ii) the conventional set
CTEQ4M (which~undershoots the CDF data at high $p_t;$ and (iii) the CTEQ4HJ
set~\cite{cteq4,cteqglu} (which contains a modified gluon distribution to
accommodate the jet data).  We see that the effect of the new quark component
is similar to that found for additional gluons at large $x.$ Thus, modification
of the quark distributions can be considered as an alternative to the large-$x$
gluon as the source of the jet excess if it persists. We note that both CDF and
D0 experiments have found that the angular distribution of these jet events is
consistent with QCD predictions
\cite{JetAngDis}. Hence, in this case, at least, it is important to have
viable Standard Model explanations of the observed cross sections.

If with more data the excess at HERA turns out to be larger than that shown in
Fig.~\ref{totqrat}, can one go further and increase the quark distributions
even more?  Two possible scenarios come to mind.  The first is to make the
additional component even more peaked towards $x=1$ than in our toy model.  We
have obtained acceptable fits under this scenario with a 1\% additional
component of $u_v$, and this effectively doubles the size of the effect shown
in Fig.~\ref{totqrat}.  Beyond 1\%, it becomes very difficult to reconcile with
the fixed target DIS measurements, especially BCDMS since it is sensitive to
the electric charge squared and we are adding charge 2/3 quarks.\footnote{One could also wonder whether uncertainties on the data in the last
$x$ bin of the BCDMS measurement might be larger then estimated.  If one allows
for a relatively small mismeasurement in this bin then it becomes much easier
to increase the effect at HERA.  We know of no reason to question this
measurement, but if it means the difference between a new physics signal or a
Standard Model explanation it should be closely scrutinized.}  The second
possibility to increase the effect at HERA beyond our toy model is to make an
additional modification of one of the charge 1/3 quark distributions, most
likely the $d$ valence distribution.  The BCDMS measurement on 
hydrogen mentioned above is four times less sensitive to charge 1/3 
quarks,  while for the HERA data Z exchange is larger than photon 
exchange and therefore 
almost equally sensitive to either electric charge.  
We note that the best measurement of this distribution is the BCDMS 
measurement with a deuterium target, and the last $x$
bin of this measurement is 20\% above NLO QCD using conventional parton
distributions.  This is shown in Fig.~\ref{bcdeu}, where the relative
difference between data and theory is shown, using both CTEQ4M and MRSR1 parton
distributions. Here, an increase in $d_v$ at large $x$ will actually improve
agreement with existing data.

For completeness we now review the other quark flavors.  The sea quarks are
significantly smaller than the valence quarks at large $x$, but also less well
measured.  There are several measurements that indirectly constrain the $\bar
u$ and $\bar d$ partons, these extend out to $x=0.65$.  The strange quark
distribution is constrained both by global fits and by dimuon neutrino
scattering experiments,  but also with a limited $x$ range.  
Overall, it seems very unlikely that any modification
of $\bar u,\bar d,s$ could ever be large enough to compare to the valence
distributions.  Another intriguing possibility is that of intrinsic heavy
flavor.  The charm and bottom distributions are normally generated completely
from perturbative gluon splitting.  Intrinsic heavy flavor would be an
additional non-perturbative component that is now assumed to be zero.  In
addition, these distributions are predicted in non-perturbative
models~\cite{brod,jack} to peak at relatively large $x$ due to their larger
mass.  The 0.5\% additional component of $u_v$ in our toy model could be
replaced with intrinsic charm with identical results.  If intrinsic charm
exists, then intrinsic bottom probably also exists with a smaller total
momentum fraction.  But since the evolution of the bottom quarks would not
start until $Q_0^2=25$ GeV$^2$, much of the bottom quark distribution simply
by-passes the fixed target region in its QCD evolution, cf. Fig.~\ref{evolve}.
The combination of this effect with the charge 1/3 advantage mentioned earlier
would make it much easier to fit fixed target and HERA at the same time,
thereby perhaps compensating for the expected smaller momentum fraction.  Even
though the sea quarks are typically much smaller than valence at these $x$
values, if the HERA excess persists there are scenarios with intrinsic heavy
flavor that should be investigated as possible modifications of the structure
of the proton.

In summary, we have investigated effects of the 
current uncertainty of the quark
distribution functions at large $x$ in the kinematic region of the HERA excess.
Using conventional parametrizations, we confirm the $6-7\%$ uncertainty quoted
by H1 and ZEUS.  However, it is possible to enhance the quark contribution to
DIS cross-sections at high-$x$, high-$Q^2$ while maintaining good agreement
with established fixed target data by introducing a small additional high $x$ component to some quark flavors, through the feed-down mechanism of QCD evolution.  The
enhanced quarks can provide an alternative explanation of the CDF
jet excess. 
Even if the scenarios described in this note eventually turn out to be 
improbable,  for either theoretical or experimental reasons, they 
do demonstrate that conventional estimates of a small Standard Model 
uncertainty for the HERA measurements rest on implicit assumptions 
of the non-perturbative quark distributions which are not founded 
on known experimental results.  Effects such as those explored 
here could also be only a part of the explanation of the observed 
events, if they persist and if they prove to be smooth in their 
$x$-dependence.

To pursue these effects further, it is important to investigate 
in more detail all processes which are sensitive to quarks 
at large $x$.
Fig.~\ref{evolve2}
illustrates the kinematic regions of three different classes of measurements
that could in the future provide additional constraints on large-$x$ quark
distributions.  They include fixed target 
data beyond $x>0.75$ that are not currently 
being used due to concerns about nuclear or 
higher twist effects.  There are also jet 
measurements at the Tevatron and other 
data from HERA that should provide useful constraints.   
In addition,  whether the large $x$ component is due to heavy
flavors needs to be investigated by measuring the flavor content of the HERA
events and the CDF high $p_t$ jets.

We thank members of the CTEQ collaboration for useful discussions.  This work 
was supported in part by the Department of Energy and the National Science 
Foundation.

\figevolve
\figparam
\figuvadd
\figtotq
\figbcd
\figjet
\figbcdeu
\figev2

\end{document}